\documentclass[twocolumn]{aastex63}
\usepackage{textcomp}
\usepackage{mathtext}
\usepackage{amssymb}
\usepackage{rotating}

\newcommand{\fermi}{\textit{Fermi}}
\newcommand{\gr}{$\gamma$-ray}
\newcommand{\tgt}{V404 Cygni}

\shorttitle{High-energy $\gamma$-ray Detection of V404 Cygni}
\shortauthors{Xing \& Wang}

\begin{document}

\title{Detection of the microquasar V404 Cygni at $\gamma$-rays revisited: 
short flaring events in quiescence}

\correspondingauthor{Yi Xing \& Zhongxiang Wang}
\email{yixing@shao.ac.cn; wangzx@ynu.edu.cn}

\author{Yi Xing}
\affiliation{Key Laboratory for Research in Galaxies and Cosmology,
Shanghai Astronomical Observatory, Chinese Academy of Sciences,
80 Nandan Road, Shanghai 200030, China}

\author{Zhongxiang Wang}
\affiliation{Department of Astronomy, School of Physics and Astronomy, 
Key Laboratory of Astroparticle Physics of Yunnan Province, Yunnan University, 
Kunming 650091, China}
\affiliation{Key Laboratory for Research in Galaxies and Cosmology,
Shanghai Astronomical Observatory, Chinese Academy of Sciences, 
80 Nandan Road, Shanghai 200030, China}

\begin{abstract}
The microquasar V404 Cygni (also known as GS 2023+338) was previously reported
to have weak GeV $\gamma$-ray emission in sub-day time periods 
during its 2015 outburst. In order to provide more detailed information 
at the high energy
range for this black hole binary system, we conduct analysis to 
the data 
obtained with the Large Area Telescope (LAT) onboard {\it the Fermi Gamma-ray
Space Telescope (Fermi)}. Both LAT database and source catalog used are 
the latest.
In addition to the previously reported detection at the peak of the 2015 
outburst,  we find possible detection ($\sim 4\sigma$) of the source during 
	3-day time period of 2015 Aug. 17--19 (at the end of the 2015 
	outburst) and one
convincing detection ($\simeq 7\sigma$) in 2016 Aug. 23--25. The  
latter high-significance detection shows
that the \gr\ emission of the source is soft with photon index 
$\Gamma\sim 2.9$.  As \gr\ emission
from microquasars is considered to be associated with their jet activity, we discuss 
the results by comparing with those well studied cases, namely Cyg X-3 and 
Cyg X-1. The detection helps identify V404 Cygni as a  
microquasar with detectable $\gamma$-ray emission in its quiescent state, and
adds interesting features to the microquasar group, or in a more general 
context to X-ray binaries with jets.

\end{abstract}

\keywords{stars: black holes --- stars: individual (V404 Cygni) --- gamma rays: stars --- X-rays: binaries}

\section{Introduction}

The X-ray binary \tgt\ (or GS~2023+338) is a stellar-mass black hole system,
consisting of a $\sim 9\ M_{\sun}$ black~hole and
a $\sim$1$\ M_{\sun}$ low-mass companion star \citep{ccn92,cc94,kfr10}.
The system has an orbital period of 6.5~days \citep{cc94} and
is at a distance of 2.39\,kpc \citep{mil+09}. Long after a previous 
outburst
in 1989 \citep{mak89}, it underwent a second outburst in 2015,
both events triggering extensive observations at multi-frequencies. 
In particular, the recent outburst, which lasted from 2015 Jun. 15
\citep{bar+15} to mid Aug. \citep{siv+15,plo+17}, was monitored 
with a variety of radio, optical, X-ray, and $\gamma$-ray facilities.
With particular attention paid to physical processes related to
the black hole accretion (see, e.g., \citealt{bm16}), 
different aspects of the binary have been learned from these observations
(e.g., \citealt{rod+15,plo+17,wal+17,tet+17,mai+17} and references therein).
In addition, an approximately month-long mini-outburst from
the source was seen at the end of 2015. This mini-outburst exhibited 
similar features to those in the main outburst, but at a lesser 
intensity (4--14 times fainter, depending on the energy range; 
see \citealt{mun+17} and references therein).

This black hole binary belongs to the microquasar category \citep{mr99}, 
as jets associated with the black hole have been observed
in its quiescence state (e.g., \citealt{gfh05,ran+16,plo+19}), as were
additional jet ejecta observed in the 2015 outburst 
(e.g., \citealt{wal+17,tet+17,tet+19}).

Microquasars are theoretically expected to be \gr--emitters
(e.g., \citealt{aa99,gak02,rom+03}).
Observationally the microquasars Cyg X-3 and Cyg X-1 were detected and
relatively well studied at
$\gamma$-rays \citep{fer09,tav+09,sab+10,bod+13,mal+13,zan+16,zdz+17}.
As for \tgt, during the second outburst,
\citet{loh+16} reported the source detection
($\sim 4.5\sigma$) above 100 MeV in a 12-h time bin 
on 2015 Jun. 26, with the data obtained
with the Large Area Telescope (LAT) onboard the {\it Fermi Gamma-ray
Space Telescope (Fermi)}. The detection was confirmed by the 
{\it AGILE} observation \citep{pia+17},
a $\sim$4.3$\sigma$ significance detection at
nearly the same time in the energy range of 60--400\,MeV.
Very-high-energy \gr\ observations were also conducted using
the MAGIC telescopes in 2015 Jun., but the source was undetected
in the 200--1250\,GeV energy range \citep{ahn+17}.

Given that \fermi\ LAT has been collecting data for nearly 12 years,
and that both the database and source catalog have been
updated several times since the previous 
work by \citet{loh+16}, we reanalyzed the \gr\ data
for \tgt, searching for new detection whilst re-examining 
previous detection results. From our detailed analyses,
additional short-term flaring events, not in the outburst but in 
quiescence, were found. 

\section{LAT Data Analysis and Results}
\label{sec:obs}

\subsection{LAT data and source model}
LAT has been scanning the whole sky continuously and collecting data in
the energy range from 50~MeV to 1~TeV since 2008 
Aug. \citep{atw+09,4fgl20}. We selected the 0.06--500 GeV LAT
events from 2008-08-04 15:43:36 (UTC) to 
2020-03-05 01:16:35 (UTC; approximately 11.5 yrs)
within a $20\arcdeg\times 20\arcdeg$ region 
centered at the position of \tgt.
The latest \fermi\ Pass 8 database was used.
Following the recommendations of the LAT team\footnote{\footnotesize http://fermi.gsfc.nasa.gov/ssc/data/analysis/scitools/}, 
the events with quality flags of `bad' and 
zenith angles larger than 90 degrees were excluded; the latter to prevent 
contamination from the Earth's limb.
\begin{figure}
\centering
\includegraphics[width=0.48\textwidth]{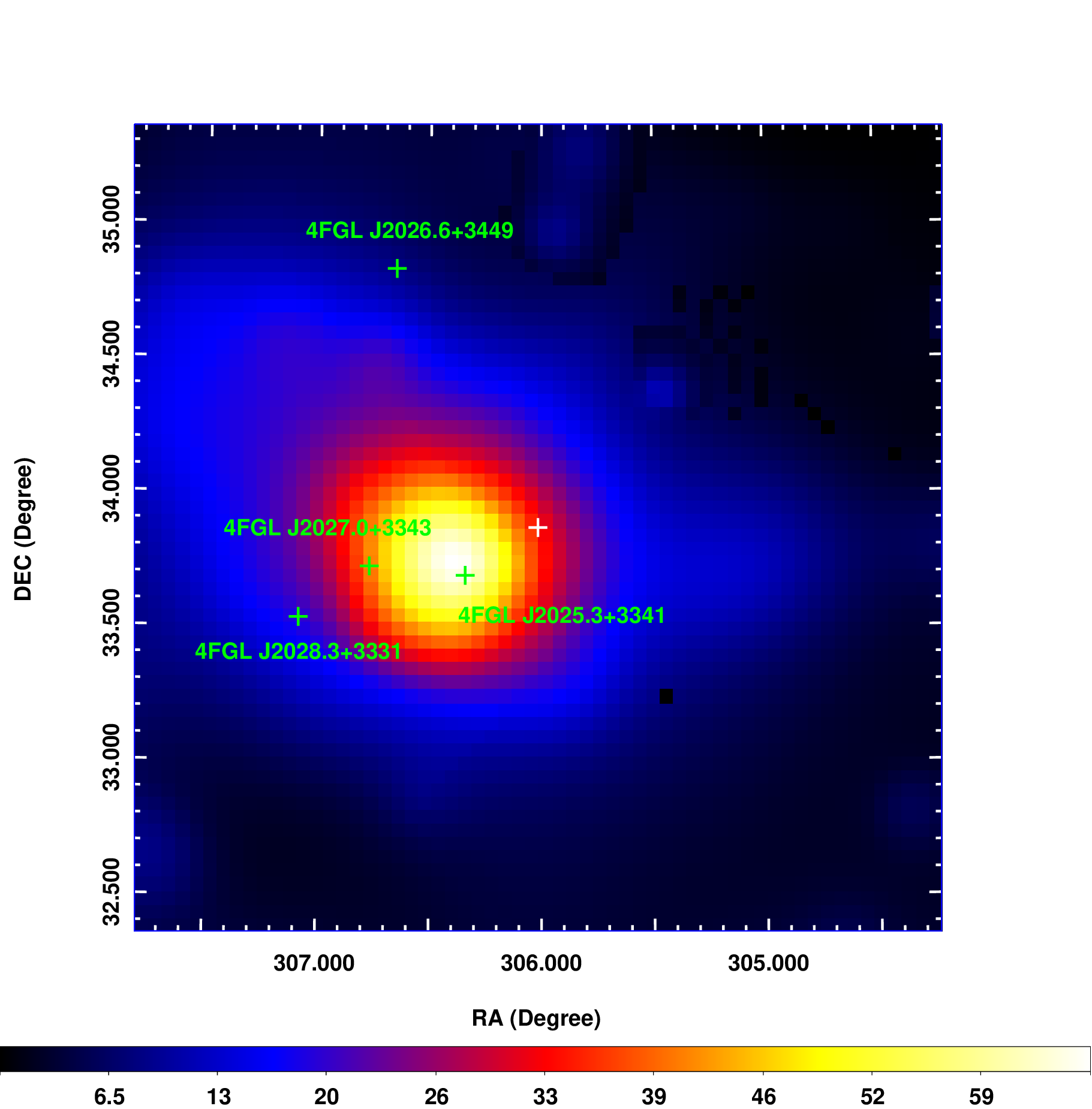}
\caption{
TS map of the $3\arcdeg\times 3\arcdeg$ region centered at \tgt\ 
in the energy range of 0.3--500 GeV from the data 
during MJD~55076--55078 (the peak time of a major flare of the blazar 
	4FGL~J2025.3+3341; see details in section~\ref{sec:nearby}).
The white plus marks the position of \tgt\  and the green pluses 
mark the nearby catalog sources. In this TS map, the other three catalog
sources are removed.
The image scale of the map is 0$\fdg05$\,pixel$^{-1}$. 
}
\label{fig:ts}
\end{figure}

Using the recently released \fermi\ LAT 10-year source catalog 
(4FGL-DR2; \citealt{bal+20}), we constructed a source model (for 4FGL, 
see \citealt{4fgl20}).
The sources listed in 4FGL-DR2 that are within a 20-degree radius of
\tgt\ were included. Their spectral forms 
are provided in the catalog.  In our analysis, 
the spectral parameters of the sources 5 degrees farther from 
the target were fixed at the values given in the catalog.
The background Galactic and extragalactic diffuse 
spectral models (gll\_iem\_v07.fits and iso\_P8R3\_SOURCE\_V2\_v1.txt
respectively) were also included in the source model, with their
normalizations set as free parameters in the analysis.

\subsection{Search for possible detection in the LAT time period}

We analyzed the 11.5 yr LAT data to search for possible \gr\ emission 
from \tgt. 
Because the instrument response function of LAT has
relatively large uncertainties in the $<$0.3~GeV energy range, and 
four catalog sources located close to \tgt\ (see below section~\ref{sec:nearby}
for details) may easily contaminate the search results due to the
large point spread function (PSF) of LAT in the low energy, we used mainly the 
0.3--500\,GeV data in our analysis.
We first performed the binned likelihood analysis to the whole selected
LAT data.  No emission was found at the position of the target,
as the obtained test statistic (TS) value was $\sim 0$.
%%(see the left panel of Figure~\ref{fig:ts}).
\begin{figure}
\centering
\includegraphics[width=0.47\textwidth]{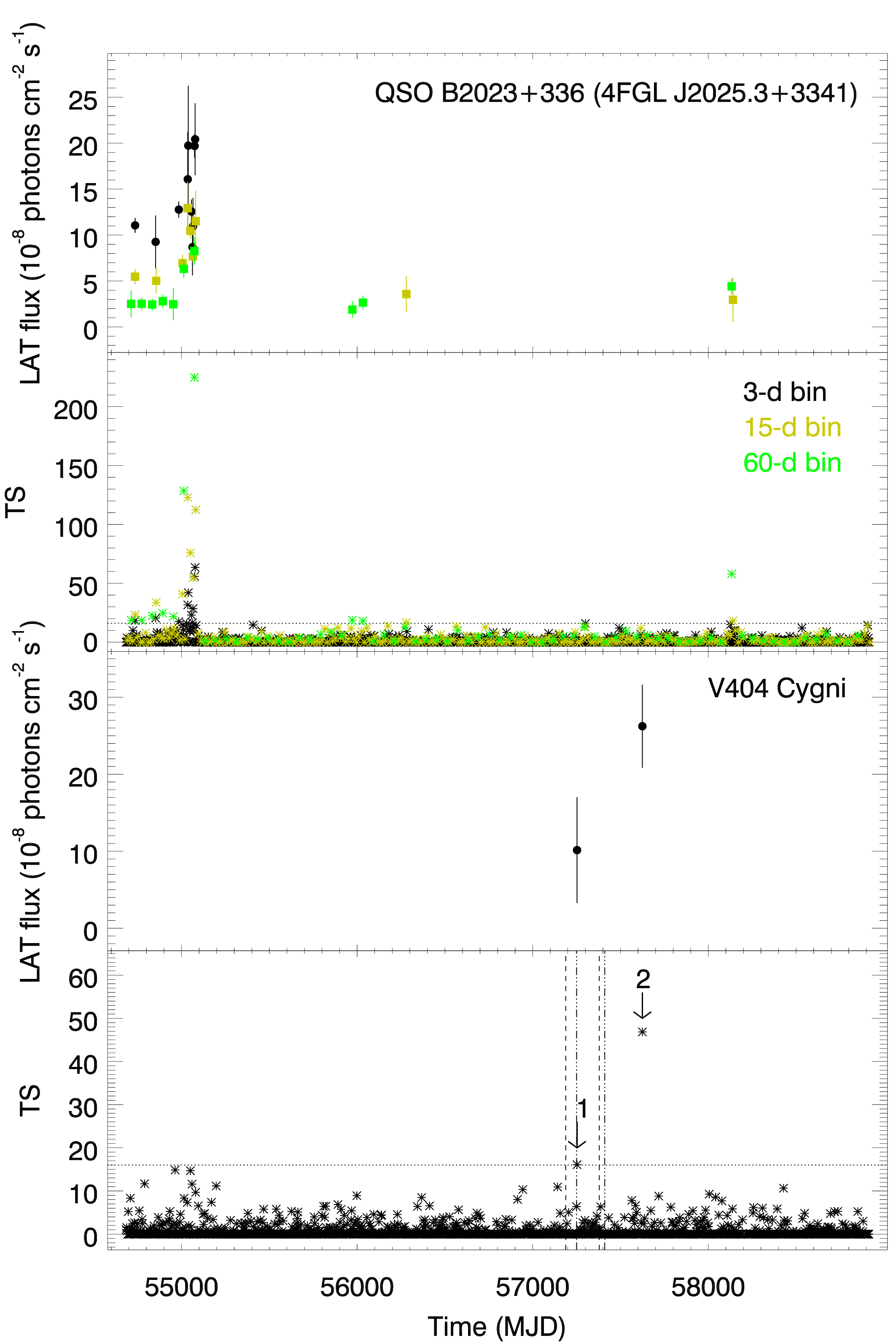}
	\caption{{\it Top two} panels: 3-day, 15-day, and 60-day binned fluxes 
	and 
	TS values (in 0.3--500\,GeV) of the blazar 4FGL~J2025.3+3341 
	(QSO B2023+336). Only those fluxes with TS$\geq$16 (marked by the 
	dotted line in the second panel) are shown in the first panel.
	The sets of fluxes illustrate a major flare 
form the blazar in the beginning of the whole time period.
	{\it Bottom two} panels: 3-day binned fluxes and TS values
	(in 0.3--500\,GeV) obtained at the position of \tgt.
	Only fluxes with TS$\geq$16 (marked by the dotted line in the 
	bottom panel) are shown in the third panel. 
Two TS$\geq$16 events, not coincident with the flares of the 4FGL~J2025.3+3341,
are marked as 1 and 2.  The time periods of the 2015 outburst
and mini-outburst are marked by the dashed (start time) and dash-dotted
	(end time) lines.} 
\label{fig:3lc}
\end{figure}

Given that both microquasars Cyg X-3 and Cyg X-1 show variable $\gamma$-ray
emission (e.g., \citealt{cor+12,bod+13}) and  were significantly detected 
in short time periods such as one day
(e.g., \citealt{sab+10,zan+16,zdz+18}),
we focused on searching for detection in different short time bins.
By testing for 1-day, 3-day, or 6.5-day 
(i.e., the orbital period of \tgt), we
found that the analysis to the data in 3-day bins 
well show possible detection over the LAT data time period.

\subsubsection{Nearby sources}
\label{sec:nearby}

There are four nearby sources listed in the 4FGL catalog, which are
4FGL~J2028.3+3331, 4FGL~J2026.6+3449, 4FGL~J2027.0+3343, and 4FGL~J2025.3+3341 
(see Figure~\ref{fig:ts}). One of them, 4FGL~J2025.3+3341, is a blazar 
(QSO~B2023$+$336) that
was found to have had a flare at the beginning of the LAT data time 
period \citep{kar+12}.
We thus first checked properties of the sources
to avoid possible contamination.
4FGL~J2028.3+3331 is a pulsar (PSR~J2028+3332) and had stable emission 
(TS$\simeq$6000 in the whole data), while 4FGL~J2026.6+3449
and 4FGL J2027.0+3343 on the other hand were faint, having TS$\simeq$59 
and $\simeq$34 respectively
in the whole LAT data. The former is a blazar and the latter an unknown source,
and neither showed any significant flux variations.

The blazar 4FGL~J2025.3+3341 
is the closest to \tgt, having a separation of only 0\fdg28.
Previously, its \gr\ properties
were studied by \citet{kar+12} using the first 31-month LAT data, 
during which time it displayed clear \gr\ variations.
We then proceeded to analyze sets of data in time bins of 3-, 15-, 
and 60-days for this source.
In Figure~\ref{fig:ts}, we show a TS map during 
MJD~55076--55078, which according to the 3-day binned data
determined to be at the peak time of a major flare.
As can be seen, the source's flare
could contaminate our search results for short flares from \tgt\ 
due to its proximity. 
Based on the light curve results obtained from the 15- and 60-day binned
data (see the top two panels of Figure~\ref{fig:3lc}),
the flare was estimated to have lasted
from the beginning of the LAT data to MJD~55103. 
After this active phase, several occasional short flaring
events occurred in the rest of the time period as well.
We performed binned analysis 
to the LAT data in both the active phase (MJD~54682--55103) as well as
the rest of the time period for this blazar. No significant spectral changes 
were found, as the photon index $\Gamma$ was 2.75$\pm0.08$ 
in the former time period and 2.77$\pm$0.04 in the latter.

\subsubsection{Search results for \tgt}
We performed unbinned likelihood analysis to 3-day binned LAT data in
0.3--500\,GeV band for \tgt.
In this analysis, because the nearby blazar 4FGL~J2025.3$+$3341 had 
variable flux, its spectral normalization was set as a free parameter, while
all other spectral parameters of the 4FGL sources within 
5 degrees from the target were fixed, at the values obtained from the above 
binned likelihood analysis to the whole data. 
In the bottom panel of Figure~\ref{fig:3lc}, the TS values obtained at 
the position of the target are shown.
\begin{figure*}
\centering
\includegraphics[width=0.42\textwidth]{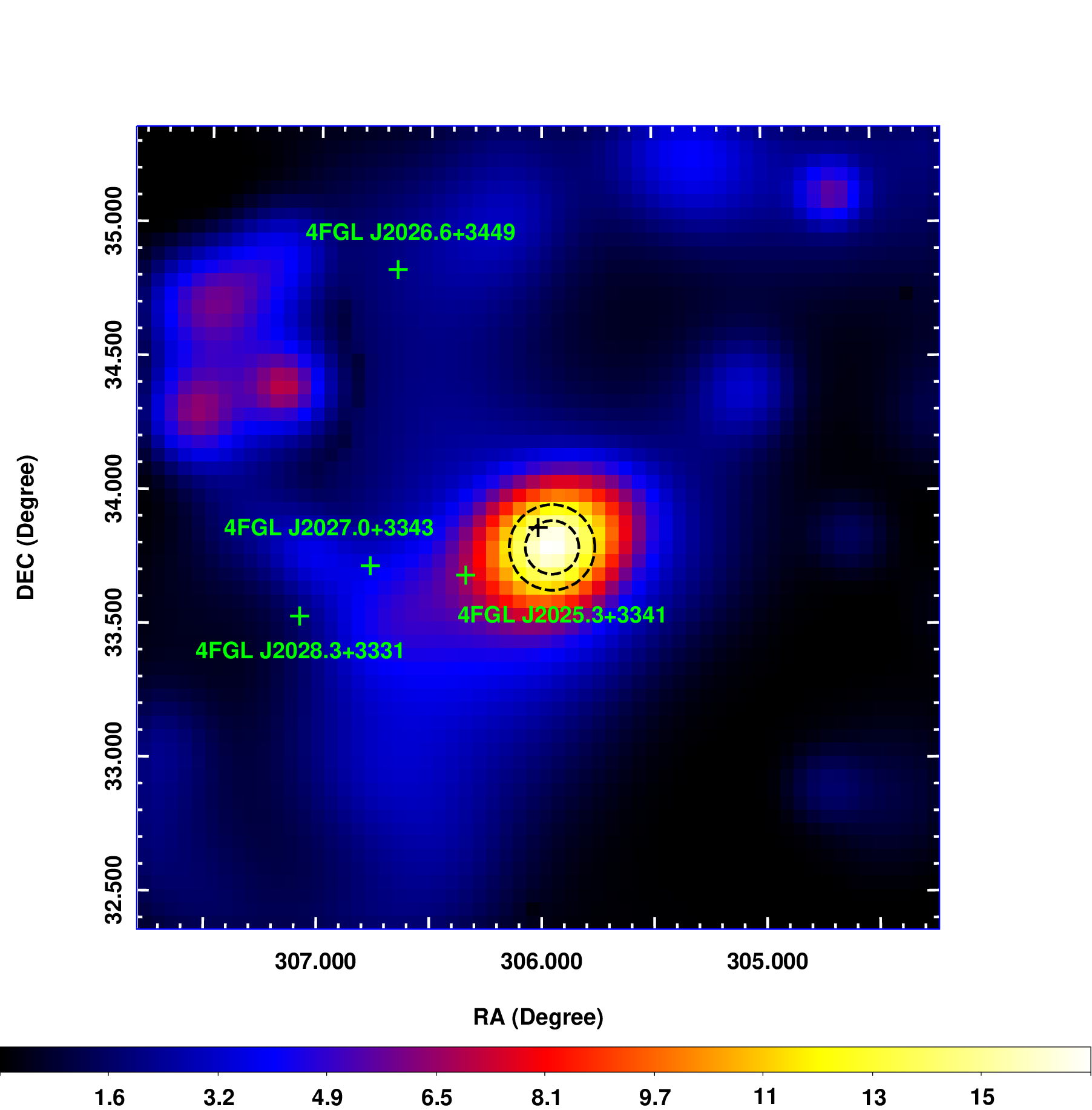}
\includegraphics[width=0.42\textwidth]{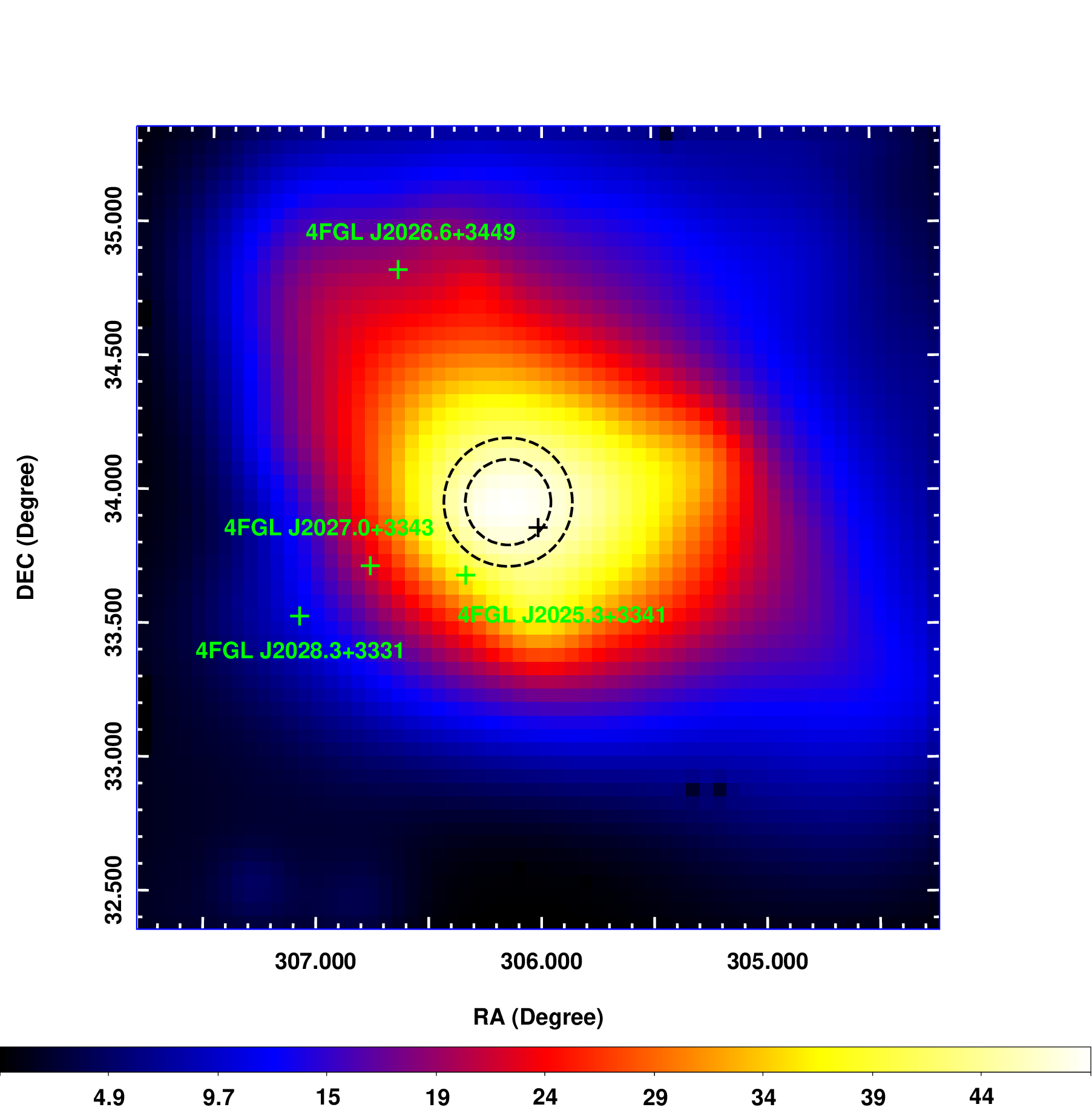}
\caption{
TS maps of the $3\arcdeg\times 3\arcdeg$ region centered at \tgt\ 
in the energy range of 0.3--500 GeV obtained from the data in MJD 57251--57253 
({\it left}) and MJD 57623--57625 ({\it right}).
The black plus marks the position of \tgt\  and the dashed circles indicate
the 1$\sigma$ and 2$\sigma$ error circles determined for the excess emission.
The green pluses mark the catalog sources that were considered in
	the source model and removed in the TS maps.
The image scale of the maps is 0$\fdg05$\,pixel$^{-1}$.}
\label{fig:p2}
\end{figure*}

The 3-day binned TS data points indicate several low-significance 
hints of 
detection (9$<$TS$<$16) of \tgt\ during the active phase of the nearby blazar. 
These hints of detection were neglected in order to avoid any contamination
due to the overlapping with the active phase.
We also analyzed the whole active phase of the blazar 4FGL~J2025.3$+$3341 
(MJD 54682--55103), and \tgt\ was not detected as a persistent 
source during this long integration time (TS$\sim$3). 
When we required TS$\geq$16 as a criterion for possible detection,
two gamma-ray events have been found: MJD 57251--57253 and 
57623--57625. The first one, on 2015 Aug. 17--19 (TS$\sim$16), occurred 
at the end of the 2015 multi-frequency outburst. The second one, on 
2016 Aug. 23--25, is significant with TS$\sim$46 and coincident 
with a quiescent multi-frequency phase of the microquasar. Since 
they did not occur during the active phase, nor during the short flaring events 
of the nearby blazar, we 
focused on these two transient events and analyzed their 
characteristics in detail.

Considering that the \gr\ event 
at the peak of the 2015 outburst, as detected by \fermi\ LAT and 
AGILE \citep{loh+16,pia+17}, is characterized by a 
soft spectrum, we also performed an unbinned likelihood analysis of 
the 3-day \fermi\ LAT light curve in the 0.06--0.3 GeV energy range. We 
found four events with TS$\geq$16, at 
MJD~55031--55033, 55037--55039, 55046--55048, and 55064--55066. However, since these events 
were observed during the major flare of the nearby blazar, possibly 
indicating a residual contamination from 4FGL~J2025.3$+$3341 in the low-energy 
range, we neglected them as possible detections of \tgt. 
As a result, no significant detections have been found in the low energy 
range, outside of the active phase of the blazar.

\subsubsection{Possible detection in MJD 57251--57253}
\label{sec:3}

The first possible detection only has TS$\simeq$16, right at the end of 
the 2015 outburst (during Aug. 17--19; Figure~\ref{fig:3lc}). 
In order to verify the detection, we calculated a TS map of
a $3\arcdeg\times 3\arcdeg$ region centered at the position of \tgt\  
in the 0.3--500~GeV energy range, shown in the left panel of 
Figure~\ref{fig:p2}.
Because the time bin does not overlap with any obvious flares
from the nearby blazar, the blazar was considered in the source model and 
removed in the TS map. Excess emission at the position 
of \tgt\ is clearly seen.

We then ran \textit{gtfindsrc} in {\tt Fermitools} to determine the position of
the excess emission. It has 1$\sigma$ nominal uncertainty of
0\fdg1, consistent with that of \tgt\ (see the left panel of
Figure~\ref{fig:p2}).
We performed unbinned likelihood analysis to the data with the fitted 
position and obtained a photon index of $\Gamma=$ 2.5$\pm$0.4 and 
a 0.3--500 GeV flux of
$F_{0.3-500}= ($1.1$\pm$0.5$) \times 10^{-7}$ photons~s$^{-1}$\,cm$^{-2}$ 
(with a TS value of 17). 

\subsubsection{Detection in MJD 57623--57625}

Since this event has TS$\simeq$46, we consider it as a likely detection
of \tgt.
We performed the same analysis to the 0.3--500 GeV data in the 3-day binned
data as mentioned in
Section~\ref{sec:3}. The TS map of the $3\arcdeg\times 3\arcdeg$ region
shown in the right panel of Figure~\ref{fig:p2} reveals excess
emission at the position of \tgt, with a maximum 
TS$\simeq$46. The determined position for the excess
emission has a nominal uncertainty of 0\fdg16, and \tgt\ is in the
error circle. The unbinned likelihood analysis gave $\Gamma= 2.9\pm$0.3 and 
$F_{0.3-500}= ($2.6$\pm$0.5$) \times 10^{-7}$ photons~s$^{-1}$\,cm$^{-2}$
(with a TS value of 48).
%%, see Table~\ref{tab:likelihood}).
\begin{figure}
\centering
\includegraphics[width=0.42\textwidth]{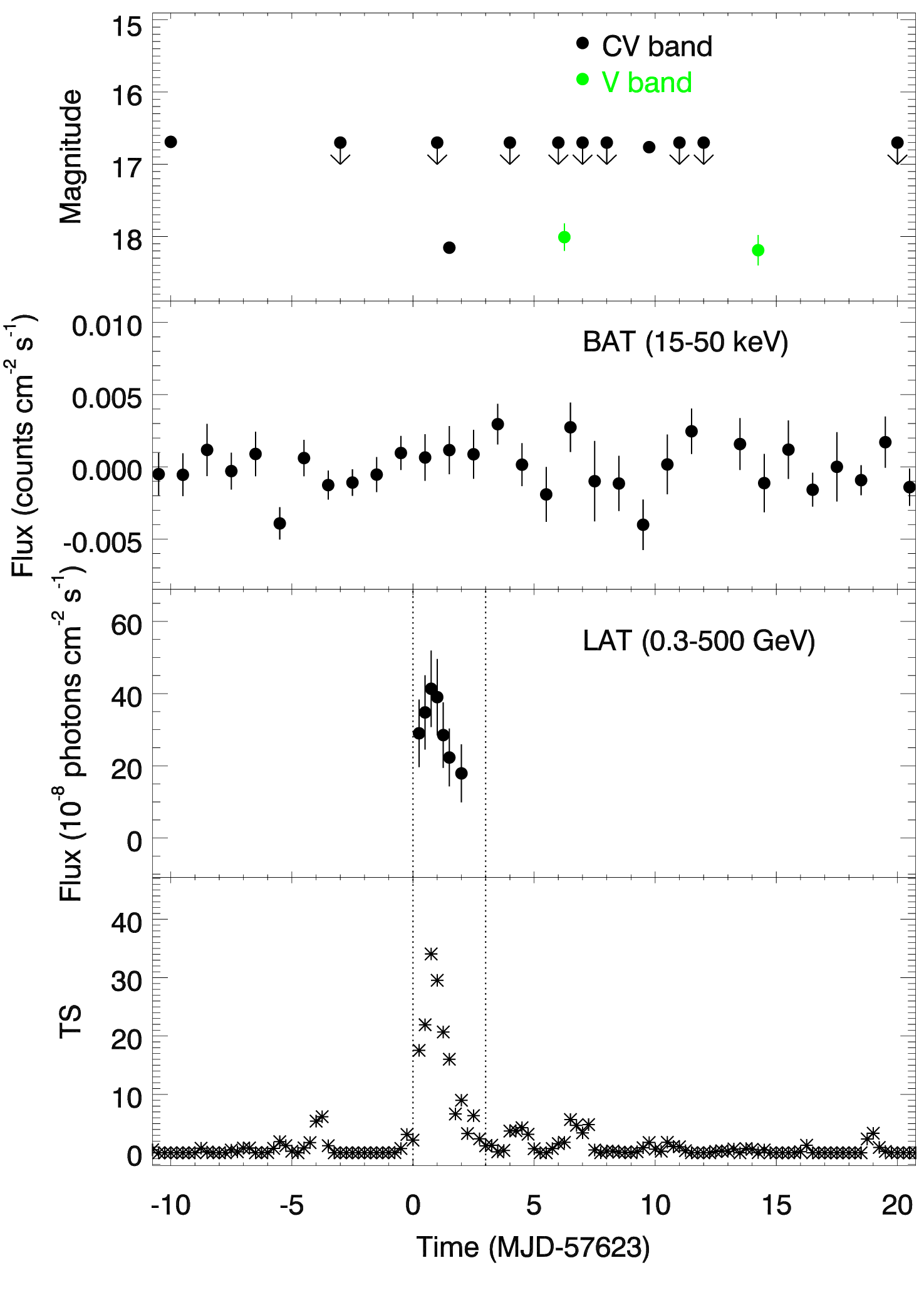}
	\caption{{\it Bottom two} panels: 1-day binned fluxes and TS values 
around the MJD 57623--57625 detection, in which each data point is shifted
	forward with 0.25~day. Only fluxes with TS$\geq$9 are shown.
	The two dotted lines indicate the time period of MJD 57623--57625.
	{\it Top two} panels: optical and X-ray monitoring data in the same 
	time period, showing that no correlated variations were seen. Among 
	the data, CV band is unfiltered photometry with V-band zeropoint.}
\label{fig:1lc}
\end{figure}

Since the excess emission has a relatively high TS value, we constructed
a short-time unbinned light curve over the detection (Figure~\ref{fig:1lc}).
In this analysis, all spectral parameters of the nearby sources, 
including the normalization of the blazar, were fixed to the values obtained 
from the preliminary binned likelihood analysis, using the data after the blazar's active phase.
We found that a light curve built upon 1-day binned data, with 0.25-day
shifts between each flux measurement,  
may uncover the more detailed variations. As shown
in Figure~\ref{fig:1lc}, the event appears to have a peak-like structure
with a duration of $\sim$2 days. 
Using the orbital parameters given in \citet{cc94}, we checked the orbital
phases for the time period.
The start and end times (MJD 57623.0 and 57626.0 respectively) 
correspond to the orbital phase 0.32 and 0.78 respectively, when the companion
star was mostly behind the black hole. 

\section{Discussion}

Having used the latest \fermi\ LAT database and source catalog, we reanalyzed
the LAT data for searching for short \gr\ events from \tgt.
We found possible detection 
during MJD~57251--57253 (with a significance of
$\sim$$4\sigma$) and another convincing one 
during MJD~57623--57625 (with a significance of $\sim$$7\sigma$).
Since the positional uncertainties determined from the \fermi\ LAT 
observations are relatively large, 
we checked the SIMBAD Astronomical Database for sources within the 
0\fdg16-radius error circle, the one derived for the convincing detection.
There are
only several known galaxies, detected by {\it NuSTAR} at the hard X-ray range of
3--24~keV and identified at optical and infrared wavelengths \citep{lan+17}.
No blazars, the dominant \gr\ sources in the sky \citep{4fgl20}, are known
to be within the error circle.

Because cases of short flares from unknown blazars were seen 
before (e.g., \citealt{lcd17,gem+19}), we tested to 
quantify the chances of coincidentally detecting a random flare. 
We randomly chose positions in 
the sky at high Galactic latitudes (to avoid the complex and 
crowded Galactic plane). At each position where there were no nearby 
$\gamma$-ray sources, we generated a 3-day binned light curve from 
the 11.5-yr LAT data in the 0.3--500~GeV energy range,
and checked to see if there was detection indicated by any
TS$\geq$25 data points. In the end, 300 such light curves were generated 
and no flares
were found (Xing \& Wang, in preparation). The results imply that the probability 
of detecting a 3-day short flare
from unknown blazars is smaller than 0.33\%. This test helps indicate that
the flare during MJD~57623--57625 is at a $\geq$3$\sigma$ confidence level 
to have not arisen from an unknown blazar.

In our search for short flaring \gr\ events from \tgt, by analyzing
the \fermi\ LAT data above 300 MeV, we did not see any significant signals 
during the main phase of the multi-wavelength 2015 outburst (Jun.\ 15--30). 
Instead, we found a low-significance event at the end of the declining 
phase of the outburst (Aug. 17--19) and a high-significant 
event (2016 Aug. 23--25) during a quiescent phase of the microquasar.
Below, we first present our more detailed 
analysis of the outburst data and the comparison of the results we obtained
with those reported in \citet{loh+16}. We also discuss the two short flares 
we have found in our search, with a focus on the more
significant one, by viewing their properties in a multi-wavelength
picture and comparing with those of \gr--emitting microquasars. In the end,
we discuss the implications of the case of \tgt.
\begin{figure}
\centering
\epsscale{1.0}
\plotone{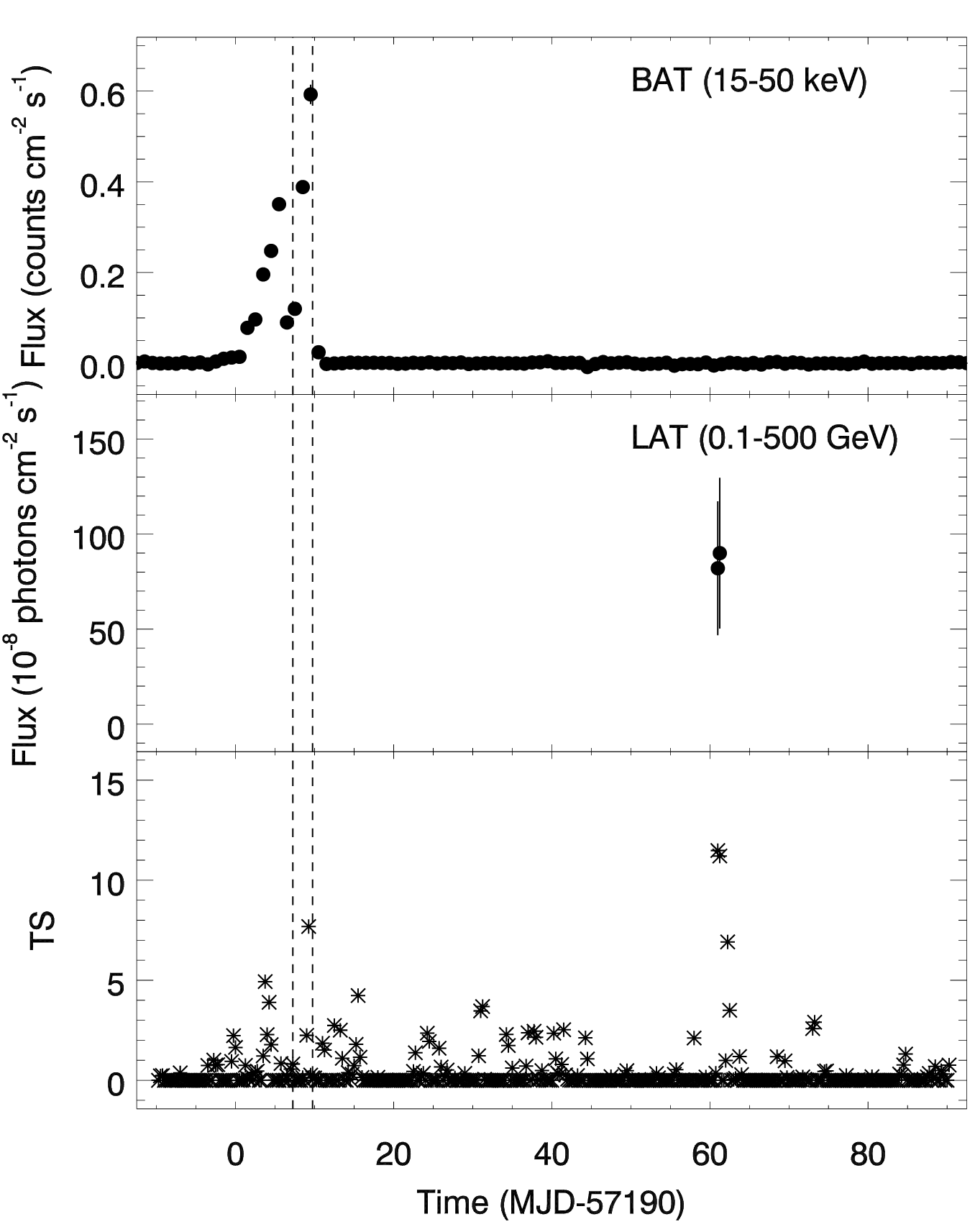}
\caption{Smoothed TS curve (12-h bin with a 6-h shift) obtained for \tgt\ 
in the energy range of 0.1--500 GeV ({\it bottom} panel). There are two TS 
points (at $\sim$MJD 57251) with values $\simeq$11, whose fluxes are shown 
	in the {\it middle} panel. 
To help indicate the outburst peak range, the {\it Swift} BAT hard X-ray
light curve is shown in the {\it top} panel. The two dotted lines mark 
the time period
of MJD 57197.25--57199.75, during which \citet{loh+16} and \citet{pia+17}
reported detection of the source with \fermi\ LAT and {\it AGILE}
respectively.}
\label{fig:olc}
\end{figure}

\subsection{2015 outburst}

In order to check for any possible detection signals during the outburst,
we followed the analysis given in \citet{loh+16}
by constructing a 12-h binned light curve, where 
each time bin was shifted by 6\,h forward (instead of 12\,h).
For this analysis, the spectral 
parameters of the 4FGL sources within 5 degrees from the target 
were firstly obtained from standard binned likelihood analysis
to the 0.1--500\,GeV LAT data after the active phase of 
the nearby blazar.
Fixing the parameter at the obtained values,
unbinned likelihood analysis was performed 
to the 0.1--500 GeV LAT data in each 12-h bins during the outburst.
Since the outburst ended sometime during
2015 mid Aug. \citep{siv+15}, we extended the analysis to Aug.
(compared to Jul. 17 as the end of data in \citealt{loh+16}; see also 
their Figure~1). The obtained TS values are shown in 
Figure~\ref{fig:olc}. We found 
a relatively high TS data point, TS$\simeq$8, at MJD 57199.25, 
and the value indicates a detection significance of $\sim2.8\sigma$. 
We calculated the 95\% flux upper limit in the 0.1--500 GeV band 
during MJD~57199.0--57199.5 and obtained 
$1.7\times 10^{-6}$~photons~s$^{-1}$\,cm$^{-2}$. This upper limit is 
comparable
to the flux of ($1.4\pm0.5$) $\times 10^{-6}$~photons~s$^{-1}$\,cm$^{-2}$ obtained
in \citet{loh+16}.
We tested to set the spectral normalizations of the sources within
5 degrees as free parameters, and nearly the same results were obtained. 
On the other hand, it can be noted that there are 
two TS$\simeq$11 data points
at MJD~57251 and 57251.25, which correspond to the possible detection event
we have found in our search above.

In \citet{loh+16}, the evtype=56
events (the three best partitions from PSF 1 to 3) 
in the 0.1--100 GeV band were used, which differs
from the total data (evtype=3) in the 0.1--500 GeV band used 
in our above analysis. 
In addition, our analysis was first run through the binned maximum
likelihood process with weights included to reflect systematic uncertainties
of the diffuse background, and then via the unbinned process without weighting
based on the model obtained from the first step. The differences could be
sensitive factors to marginal detection cases, causing inconsistent results.

Nevertheless, in our analysis of the Pass 8 data including the latest 
catalog sources, we obtained a marginal detection ($\sim$2.8$\sigma$) 
at MJD 57199.0--57199.5, consistent with the time interval 
of the low-significance flare reported in \citet{loh+16},
for a similar set of data parameters.
Although the detection significances are relatively 
low, the contemporaneous observations by \fermi\ LAT and AGILE of 
\gr\ emission at the peak of the multi-wavelength outburst 
(radio, X-ray and soft $\gamma$-ray) give statistical robustness 
to this high-energy transient event.

\subsection{Properties of the short \gr\ flares}

Both short flaring events found in our search have soft emission, 
as the obtained photon indices 
are $\Gamma\sim$ 2.5 and 2.9 respectively.
We note that
the emission reported at the peak of the 2015 outburst was even softer
($\Gamma \sim 3.5$). The softness of the source's emission is
similar to that observed in the flaring events of Cyg X-3 and Cyg X-1
\citep{fer09,zdz+18,zdz+17}. In addition, \gr\ emission is seen at spots
in the lobes of the jets from the microquasar SS~433 \citep{abe+18,xin+19},
and it is extremely soft with $\Gamma\sim 6$
in the energy range of $\leq$1.8~GeV \citep{xin+19}. These 
similarities support the association of the flaring events with \tgt, and
moreover suggest a similar mechanism for \gr\ emission from microquasars.

Different models with consideration of jets have been proposed 
to explain \gr\ 
emission from microquasars (e.g., \citealt{aa99,gak02,rom+03}). 
Based on detailed studies of Cyg X-1 (also Cyg X-3; see, e.g., 
\citealt{zdz+18}), which are able to fit its broad-band spectrum from radio 
to $\gamma$-rays, the \gr\
emission is determined to likely contain the components due to 
the jets' synchrotron self-Compton radiation
and the upscattering of photons from the accretion disk and companion star
\citep{zdz+14,zdz+17}. For \tgt, the same scenario may be at work to
give rise to its \gr\ emission.
However, different from Cyg X-1 and Cyg X-3, the densities of photons 
from the companion star of \tgt\ would be much lower, leading to
the possibility that the photons may arise mainly from the jets themselves,
and thus that the jets' synchrotron self-Compton radiation may be the 
dominant \gr\ emission mechanism.

At the peak of the outburst, \tgt\ was observed to have discrete 
jet ejecta whose emission is dominant at wavelengths from radio to 
sub-millimeter (sub-mm; \citealt{tet+17,tet+19}). 
Thus, the short transient emission detected by \fermi\ LAT and AGILE at 
the peak of the multifrequency outburst \citep{loh+16,pia+17} could arise 
from these plasma ejecta along the jets.
According to radio and sub-mm
observations, the compact jets in the system became dominant over the ejecta 
several days after the peak.
Therefore, the \gr\ events---occurring during the quiescence phase---
could be associated with
the compact and persistent jets of this microquasar.
By referring to the 
\gr\ event of 2016 Aug. 23--25 (MJD 57623--57625),
the observed 0.3--500 GeV (isotropic) luminosity is 
$\sim 1.8\times 10^{35}$\,erg\,s$^{-1}$. The jet luminosity 
(collimation corrected; see \citealt{lkp17})
from \tgt\ could be a factor of $\sim$20 larger, that is 
$\sim 4\times 10^{36}$\,erg\,s$^{-1}$. 
The mass accretion rate of the binary was estimated to be 
$\sim 7\times 10^{16}$\,g\,s$^{-1}$ \citep{mm06}, taken from 
fitting the optical and infrared broad-band spectrum of the source in 
quiescence.  The rate indicates an accretion power of
$\sim 6\times 10^{37}\eta$\,erg\,s$^{-1}$ (where $\eta$ is the efficiency)
in the system, suggesting that there is sufficient energy to provide 
the required power for the jets.

However, no variations correlated with the short flares are seen in
X-ray and optical monitoring of \tgt,
although we note that
both the {\it Swift}/BAT \citep{kri+13} and MAXI \citep{maxi09} 
do not provide sufficiently sensitive monitoring of the source 
at X-rays.
In Figure~\ref{fig:1lc}, we show the brightness measurements or upper limits
from {\it Swift}/BAT and 
the AAVSO database\footnote{\footnotesize https://www.aavso.org/} over
the MJD~57623--57625 event.
No signs of significant brightening events around the flare
were seen in the X-ray and optical data.
Unfortunately, no simultaneous radio observations of the
source are available, as it has been clearly shown 
in Cyg~X-3 that the source's \gr\ flares are due to jet activity, indicated
by correlated radio variations \citep{cor+12}. 
In any case, \tgt\ in quiescence has shown both 
significant long-term and short-term radio flux variations \citep{plo+19}. 
In the latter cases,
the variations were seen to exhibit a possible pattern of 
fast rise and slow decay.  The \gr\ flare in MJD 57623--57625 may be slightly
different, having a sharp peak-like shape (Figure~\ref{fig:1lc}) with
the timescale ($\sim 2$\,days) much longer than hour-long variation 
timescales observed at radio. 

\subsection{Implications}

The detection of the \gr\ flaring events from \tgt\ in its quiescent state
helps identify the source as another microquasar with 
detectable \gr\ emission, although the events were seemingly occasional.
The event in 
MJD 57251--57253, which occurred during the descending phase of the 
2015 multi-frequency outburst, might provide a hint of high-energy 
emission related to discrete relativistic jet ejecta, observed 
at radio and sub-mm wavelengths or indirectly derived from X-ray 
observations \citep{tet+17,tet+19,wal+17}. 
On the other hand, the phenomenology of the event detected in 
MJD 57623--57625 would suggest a different mechanism. In the context 
of a leptonic emission mechanism (upscattering of soft photons by 
relativistic jet particles), since no simultaneous optical enhancements 
were observed (see Figure~\ref{fig:1lc}), we can suppose a short increase of 
the high-energy particle rate along the compact jets.
In order to identify how the high-energy emission is related to jet activity 
in this case,
simultaneous $\gamma$-ray and radio detection will provide key information. 
However it is not easy to obtain such detection 
since based on our search, detectable \gr\ flaring events
from \tgt\ are rare. 

It is notable that different from Cyg X-3 and Cyg X-1 (as SS~433
is a peculiar case; \citealt{abe+18}) that have a high-mass companion, 
\tgt\ belongs to the more general low-mass X-ray binary (LMXB) class. 
We thus may consider that LMXBs with jets (including neutron star LMXBs;
e.g., \citealt{rus+06}) might all be able to produce 
some sorts of \gr\ emission, and searches
for short-term flaring events from them might produce interesting results.

\acknowledgements
We are grateful for constructive suggestions from anonymous referees.
We thank F. Xie for useful discussion about jet activity in different 
black-hole X-ray binaries.
This research made use of the High Performance Computing Resource in the Core
Facility for Advanced Research Computing at Shanghai Astronomical Observatory.
This research was supported by the National Program on Key Research 
and Development Project (Grant No. 2016YFA0400804) and
the National Natural Science Foundation
of China (11633007, U1738131). Z.W.  acknowledges the support by the Original 
Innovation Program of the Chinese Academy of Sciences (E085021002).

\end{document}